\documentclass[
12pt,
preprint,preprintnumbers,nofootinbib,
groupedaddress,superscriptaddress,amsmath,amssymb]{revtex4}
\usepackage{graphicx}
\usepackage{dcolumn}
\usepackage{bm}
\usepackage{amssymb}
\usepackage{amsmath}
\usepackage{epsfig}    
\usepackage{color}
\usepackage{slashed}
\usepackage{hhline}

\def\be{\begin{equation}}
\def\ee{\end{equation}}
\newcommand{\bea}{\begin{eqnarray}}
\newcommand{\eea}{\end{eqnarray}}
\newcommand{\nn}{\nonumber}

\numberwithin{equation}{section}

\begin{document}

{\begin{flushright}{KIAS-P17031}
\end{flushright}}

\title{Loop induced type-II seesaw model and GeV dark matter \\ with\\ $U(1)_{B-L}$ gauge symmetry }
%

\author{Takaaki Nomura}
\email{nomura@kias.re.kr}
\affiliation{School of Physics, KIAS, Seoul 02455, Korea}

\author{Hiroshi Okada}
\email{macokada3hiroshi@cts.nthu.edu.tw}
\affiliation{Physics Division, National Center for Theoretical Sciences, Hsinchu, Taiwan 300}

\date{\today}

\begin{abstract}
We  propose a model with $U(1)_{B-L}$ gauge symmetry and several new fermions in no conflict with anomaly cancellation
where the neutrino masses are given by the vacuum expectation value of Higgs triplet induced at the one-loop level.
The new fermions are odd under discrete $Z_2$ symmetry and the lightest one becomes dark matter candidate.
We find that the mass of dark matter is typically $\mathcal{O}(1)$-$\mathcal{O}(10)$ GeV. Then relic density of the dark matter is discussed.
\end{abstract}
\maketitle
\newpage

\section{Introduction}
Radiative neutrino mass models are widely renown in connecting the neutrino masses and a dark matter (DM) candidate at low energy theory,
which are also applied to accommodate some experimental anomalies that cannot be explained by the standard model (SM).
Thus many authors have historically been working along these ideas; for example,
refs.~\cite{a-zee, Cheng-Li, Pilaftsis:1991ug, Ma:2006km, Gu:2007ug, Sahu:2008aw, Gu:2008zf, AristizabalSierra:2006ri, Bouchand:2012dx, McDonald:2013hsa, Ma:2014cfa, Kajiyama:2013sza, Kanemura:2011vm, Kanemura:2011jj, Kanemura:2011mw, Schmidt:2012yg, Kanemura:2012rj, Farzan:2012sa, Kumericki:2012bf, Kumericki:2012bh, Ma:2012if, Gil:2012ya, Okada:2012np, Hehn:2012kz, Dev:2012sg, Kajiyama:2012xg, Toma:2013zsa, Kanemura:2013qva, Law:2013saa, Baek:2014qwa, Kanemura:2014rpa, Fraser:2014yha, Vicente:2014wga, Baek:2015mna, Merle:2015gea, Restrepo:2015ura, Merle:2015ica, Wang:2015saa, Ahn:2012cg, Ma:2012ez, Hernandez:2013dta, Ma:2014eka, Ma:2014yka, Ma:2015pma, Ma:2013mga, radlepton1, Okada:2014nsa, Brdar:2013iea, Okada:2015kkj, Bonnet:2012kz, Joaquim:2014gba, Davoudiasl:2014pya, Lindner:2014oea, Okada:2014nea, Mambrini:2015sia, Boucenna:2014zba, Ahriche:2016acx, Fraser:2015mhb, Fraser:2015zed, Adhikari:2015woo, Okada:2015vwh, Ibarra:2016dlb, Arbelaez:2016mhg, Ahriche:2016rgf, Lu:2016ucn, Kownacki:2016hpm, Ahriche:2016cio, Ahriche:2016ixu, Ma:2016nnn, Nomura:2016jnl, Hagedorn:2016dze, Antipin:2016awv, Nomura:2016emz, Gu:2016ghu, Guo:2016dzl, Hernandez:2015hrt, Megrelidze:2016fcs, Cheung:2016fjo, Seto:2016pks, Lu:2016dbc, Hessler:2016kwm, Okada:2015bxa,
Ko:2017quv, Ko:2017yrd, Lee:2017ekw, Antipin:2017wiz, Borah:2017dqx, Chiang:2017tai, Kitabayashi:2017sjz, Das:2017ski, Nomura:2017ezy} mainly focusses on the neutrino mass scenarios realized at one-loop level, and 
refs. \cite{Wang:2016lve, Guo:2017ybk, Lindner:2016bgg} discuss the systematic analysis of (Dirac) neutrino oscillation, charged lepton flavor violation, and collider physics in the framework of neutrinophilic and inert two Higgs doublet model (THDM), respectively.

A Higgs triplet model (HTM) is also an interesting scenario to get non-zero neutrino masses where these masses are induced by the vacuum expectation value (VEV) of an $SU(2)_L$ triplet Higgs field, $\Delta$, and this scenario is also called as type-II seesaw model~\cite{Magg:1980ut,Konetschny:1977bn}. 
 The VEV of the Higgs triplet, $v_\Delta$, is required to be as small as $v_\Delta \lesssim 3$ GeV from the electroweak precision measurement; especially that of $\rho$-parameter.
In the Higgs triplet model, we consider the triplet $\Delta$ has electroweak scale mass term $\mu_\Delta^2 {\rm Tr}[\Delta^\dagger \Delta]$ which does not induce VEV of the triplet due to positive $\mu_\Delta^2$ in contrast to the case of SM Higgs. The VEV of Higgs triplet is induced via $H^T \Delta H$ interaction with coupling $\mu$ as $v_\Delta \propto \mu v^2/m_\Delta^2$ where $v$ is the VEV of the SM Higgs field and $\mu$ is assumed to be smaller than electroweak scale to get a small value of $v_\Delta$. 
However the value of parameter $\mu$ is not theoretically restricted and can be large.

 In ref.~\cite{Kanemura:2012rj}, the authors have introduced a model that theoretically realizes a small value of $v_\Delta$.
In this scenario,  the $H^T \Delta H$ interaction is forbidden at tree level by global $B-L$ symmetry but it is allowed by an one-loop effect where lepton number violation is included and scalar bosons with $Z_2$ odd parity propagate inside the loop. 

It is also interesting to construct such a model with $U(1)_{B-L}$ {\it gauge} symmetry since it leads rich phenomenology. In this case, we need to add several SM singlet fermions with $B-L$ charge to cancel gauge anomalies, and it would be motivated to generate the tiny $v_\Delta$ by a loop diagram containing propagators of fermions with $Z_2$ odd parity which can be a DM candidate. In addition, we would have some predictions for DM mass when $Z_2$ odd neutral fermion masses are related to $v$ and $v_\Delta$.

In this paper, we propose a neutrino model with Higgs triplet field where $v_\Delta$ is arisen at the one-loop level and we interpret this mechanism as a theoretical reason why $v_\Delta$ is so small. 
To achieve the mechanism, we impose $U(1)_{B-L}$ {\it gauge} symmetry to forbid $H^T \Delta H$ interaction at tree level and introduce several new fermions instead of scalar fields
in no conflict with anomaly cancellation.
Then $v_\Delta$ is induced via one-loop diagram with new fermion propagators after spontaneous breaking of gauge symmetries. 
Moreover, we find the typical mass scale of DM is around $1-10$ GeV,
since it is proportional to $v_\Delta$ due to a specific structure of the neutral fermion mass matrix.
Then we show the mechanism to generate the neutrino masses and analyze such a tiny mass of the fermionic DM candidate.

This paper is organized as follows.
In Sec.~II, we show our model, 
and formulated the neutral fermion sector, boson sector, lepton sector, and dark matter sector.
Also we analyze the relic density of DM without conflict of direct detection searches.
Finally We conclude and discuss in Sec.~III.


 \begin{widetext}
\begin{center} 
\begin{table}[t]
\begin{tabular}{|c||c|c||c|c|c|c|}\hline\hline  
&\multicolumn{2}{c||}{SM leptons} & \multicolumn{3}{c|}{Exotic fermions} \\\hline
Fermions& ~$L_{L_a}$~ & ~$e_{R_a}$~ & ~$L'$ ~ & ~$N_{R_a}$~ & ~$S_{R_i}$~ 
\\\hline 
 $SU(2)_L$ & $\bm{2}$  & $\bm{1}$  & $\bm{2}$ & $\bm{1}$ & $\bm{1}$   \\\hline 
$U(1)_Y$ & $-\frac12$ & $-1$  & $-\frac{1}{2}$ & $0$  & $0$    \\\hline
 $U(1)_{B-L}$ & $-1$ & $-1$  & $-1$ & $-1$  & $0$  \\\hline
$Z_2$ & $+$ & $+$  & $-$ & $-$ & $-$  \\\hline
\end{tabular}
\caption{Field contents of fermions
and their charge assignments under $SU(2)_L\times U(1)_Y\times U(1)_{B-L}\times Z_2$, where each of the flavor index is defined as $a\equiv 1-3$ and $i=1,2$.}
\label{tab:1}
\end{table}
\end{center}
\end{widetext}

\begin{table}[t]
\centering {\fontsize{10}{12}
\begin{tabular}{|c||c|c|c|c|}\hline\hline
  Bosons  &~ $H$  &~ $\Delta$  ~ &~ $\varphi$~ \\\hline
$SU(2)_L$ & $\bm{2}$ & $\bm{3}$  & $\bm{1}$  \\\hline 
$U(1)_Y$ & $\frac12$ & $1$  & $0$    \\\hline
 $U(1)_{B-L}$ & $0$ & $2$& $1$  \\\hline
$Z_2$ & $+$ & $+$& $+$ \\\hline
\end{tabular}%
} 
\caption{Boson sector, where all the bosons are $SU(3)_C$ singlet. }
\label{tab:2}
\end{table}

\section{ Model setup and phenomenologies}
In this section, we show our model and discuss some phenomenologies such as neutrino mass generation, dark matter and implications to collider physics.
First of all, we impose an additional $U(1)_{B-L}$ gauge symmetry, and introduce a vector-like fermion $L'$ with $SU(2)_L$ doublet, two right-handed neutral fermions $S_{R_i}$
\footnote{Two $S_R$ are needed to evade the massless neutral fermion that is arisen from $N_R$.},
and three right-handed neutral fermions $N_{R_a}$ with $-1$ charge under $U(1)_{B-L}$ symmetry.
Here the fermion contents and their assignments are summarized in Table~\ref{tab:1}, where $i=1,2$ and  $a=1-3$ represent the number of family.

In the scalar sector, we introduce a $SU(2)_L$ triplet field $\Delta$ which has $2$ and $1$ charges under $U(1)_Y$ and $U(1)_{B-L}$ gauge symmetry respectively.
$H$ is supposed to be the SM-like Higgs doublet with a VEV denoted by $\langle H\rangle\equiv v/\sqrt2$, while $\varphi$ is an additional Higgs singlet with nonzero VEV, $\langle \varphi\rangle\equiv v'/\sqrt2$, to realize the spontaneous breaking of $U(1)_{B-L}$ symmetry. Notice here that $\Delta$ does not have VEV at the tree level, but it is induced at the one-loop level via a diagram with propagators of the exotic neutral fermions as we discuss below. Thus its small VEV, $\langle \Delta\rangle\equiv v_\Delta/\sqrt2$, can naturally be realized. All of the scalar contents and their assignments are summarized in Table~\ref{tab:2}.
 In addition, the lightest state of these neutral fermions can be a DM candidate.
We also note that massive $Z'$ boson appears after $U(1)_{B-L}$ symmetry breaking as the other $U(1)_{B-L}$ models. 
Here $Z'$ mass is assumed to be $m_{Z'} \geq 4$ TeV to avoid the constraints from the LHC experiments when the value of $U(1)_{B-L}$ gauge coupling is $g_{BL} \sim 0.3$ as the SM $U(1)_Y$ gauge coupling.
In this paper, we briefly discuss of $U(1)_{B-L}$ breaking below and abbreviate details of the gauge interactions because it is almost the same as the others. 

\subsection{Yukawa interactions and scalar sector}
{\it Yukawa Lagrangian}:
Under our fields and symmetries, the renormalizable Lagrangians for quark and lepton sector are given by 
\begin{align}
-{\cal L}_{L}&=
(y_\ell)_{a b}\bar L_{L_a} e_{R_b} H+(y_\nu)_{ab}\bar L_{L_a} \tilde\Delta^* L_{L_b}
 + y_{N_b} \bar L'_L \tilde H N_{R_b} 
+y_{S_{aj}} \bar N^C_{R_a}  S_{R_j} \varphi\nn\\
&+ y_{\Delta_L} \bar L'^C_L \tilde\Delta^* L'_L
+ y_{\Delta_R} \bar L'^C_R \tilde\Delta^* L'_R
+ M_L \bar L'_L L'_R + M_{S_{ij}} \bar S^C_{R_i} S_{R_j} 
+{\rm c.c.},
\label{eq:lag-lep}
\end{align}
where $\tilde H(\tilde\Delta) \equiv (i \sigma_2) H^*(\Delta^*)$ with $\sigma_2$ being the second Pauli matrix, $(a,b)$ runs over $1$ to $3$, and $(i,j)$ runs over $1$ to $2$.

\subsection{Fermion Sector}
First of all we define the exotic fermion as follows:
\begin{align}
L'_{L(R)}\equiv 
\left[
\begin{array}{c}
N'\\
E'^-
\end{array}\right]_{L(R)}.
\end{align}
Then the mass eigenvalue of charged fermion $E'^\pm$ is straightforwardly given by $M_L$ in Eq.(\ref{eq:lag-lep}).
The mass matrix for the neutral exotic fermions is seven by seven in basis of $\Psi\equiv [S_{R_j},N_{R_b},N'_R,N'^C_L]^T$, and given by
\begin{align}
M_N(\Psi)=
\left[\begin{array}{cccc}
(M_S)_{ij} & m_{NS_{ib}}^T & 0 & 0  \\
m_{NS_{aj}} & {\bf 0}_{ab} & 0 & m_{NN'_a}^T \\
0 & 0 & m_R & M_L \\
0 & m_{NN'_b} & M_L &  m_L \\
\end{array}\right],
\label{eq-Nmass}
\end{align}
where $(i,j)=1,2$, $a=1\sim3$, $m_{L(R)}\equiv y_{\Delta_{L(R)}} v_\Delta/\sqrt2$, $m_{NN'_a}\equiv y_{N_a} v/\sqrt2$,  $(m_{NS})_{ij}\equiv y_{S_{ij}} v'/\sqrt2$. Then this matrix can be diagonalized by seven by seven orthogonal  matrix $O$ as $m_{\psi_i}\equiv (O M_N O^T)_i$ ($i=1\sim7$), where $m_{\psi_i}$ indicates the mass eigenvalue. Moreover we define $m_{\psi_1}\equiv m_X$ which is the lightest mass eigenvalue and $\psi_1 \equiv X$ is a DM candidate. We will discuss relic density of the DM candidate below.


\subsection{ Scalar potential}
The renormalizable scalar potential is given by
{\begin{align}
V = &
-\mu_{\varphi}^2 |\varphi|^2 - \mu^2_H |H|^2 + \mu_\Delta^2 {\rm Tr}[\Delta^\dag\Delta]  + \lambda_\varphi |\varphi|^4 + \lambda_H |H^\dag H|^2+ \lambda_{\Delta} ({\rm Tr}[\Delta^\dag\Delta])^2 + \lambda'_{\Delta} {\rm Det}[\Delta^\dag\Delta]\nn\\
+&\lambda_{\varphi H} |\varphi|^2 |H|^2  +\lambda_{\varphi \Delta} |\varphi|^2  {\rm Tr}[\Delta^\dag\Delta]
 +\lambda_{H \Delta} |H|^2 {\rm Tr}[\Delta^\dag\Delta] +\lambda'_{H \Delta} (H^\dag \sigma_i H) {\rm Tr}[\Delta^\dag\sigma_i\Delta]
\label{eq:lag-pot-2},
\end{align}
where we choose $\mu_{\varphi,H,\Delta}^2 > 0$ in the potential so that $\langle \Delta \rangle \equiv v_\Delta/\sqrt{2}=0$ at the tree level. 
 On the other hand, the VEVs of $H$ and $\varphi$, $\langle H (\varphi) \rangle = v(v')/\sqrt{2}$ , are obtained by inserting the tadpole conditions $\partial \langle V \rangle/\partial v(v')=0$, and their forms are given by
\begin{equation}
v = \sqrt{\frac{\mu_H^2 + \lambda_{\varphi H} v'^2/2 }{\lambda_H}}, \quad v' = \sqrt{\frac{\mu_\varphi^2 + \lambda_{\varphi H} v^2/2 }{\lambda_\varphi}}.
\label{eq:VEVs}
\end{equation}
Then the $U(1)_{B-L}$ is spontaneously broken by nonzero VEV of $\varphi$.
Note that since we consider $m_{Z'} \geq 4$ TeV and $g_{BL} \sim 0.3$ the $v'$ is assumed to be $v' \gtrsim \mathcal{O}(10)$ TeV; here the mass of $Z'$ is given by $m_{Z'} = g_{BL} v'$ after the symmetry breaking. 
After the spontaneous $U(1)_{B-L}$ symmetry breaking, an effective interaction term $\mu H\tilde\Delta^T H$ is given via one-loop diagram in Fig.~\ref{fig:VEV}, and $\mu$ is given by
\begin{align}
\mu &=\frac{3 y_\Delta y_{N_a} (m_{NS}^\dag)_{ai} M_{S_i}^{-1} (m_{NS}^*)_{ib}  y_{N_b}}{(4\pi)^2}
\int[dX_4] \frac{x_1}{x_1 r_L + x_2 r_{N_a}+ x_2 r_{N_b} +x_4},
\label{eq:mu}
\end{align}
where $[dX_4] \equiv dx_1dx_2 x_3x_4\delta(1-x_1-x_2-x_3-x_4)$, $r_f\equiv \frac{M_f^2}{M_{S_i}^2}$. 
 Approximating the loop integration by $\mathcal{O}(1)$ constant, the typical value of $\mu$ is roughly estimated by
\begin{equation}
\mu \sim 20 \left( \frac{m_{NS}}{\rm TeV} \right)^2 \left( \frac{\rm TeV}{M_S} \right) y_\Delta y_N^2 \ [{\rm GeV}]
\label{eq:mu}
\end{equation}
where we have omitted flavor indices for simplicity.  We thus find $\mu = \mathcal{O}(10)$ GeV when the Yukawa couplings have $\mathcal{O}(1)$ values and mass parameters $M_S(M_{NS})$ are TeV scale; this scale of $M_{NS}$ is natural when we take a scale of $U(1)_{B-L}$ breaking VEV as $\mathcal{O}(10)$ TeV while $M_S$ can be larger since it is free parameter and $\mu$  becomes smaller as it becomes larger. Thus we can obtain small $\mu$ by choosing small Yukawa couplings and/or large $M_S$.
The resulting scalar potential in the HTM potential is given by 
\begin{align}
V_{HTM}=&-\mu H\tilde\Delta^T H + \mu^2_H |H|^2 + \mu_\Delta^2 {\rm Tr}[\Delta^\dag\Delta] + \lambda_H |H^\dag H|^2+ \lambda_{\Delta} ({\rm Tr}[\Delta^\dag\Delta])^2 + \lambda'_{\Delta} {\rm Det}[\Delta^\dag\Delta]\nn\\
+& \lambda_{H \Delta} |H|^2 {\rm Tr}[\Delta^\dag\Delta] +\lambda'_{H \Delta} (H^\dag \sigma_i H) {\rm Tr}[\Delta^\dag\sigma_i\Delta]
 \label{eq:lag-effpot},
\end{align} 
where we assume $\mu$ is positive to get $v_\Delta >0$.
The triplet $v_\Delta$ is proportional to $\mu$~\cite{Kanemura:2012rj, Okada:2015nca};
\begin{align}
v_\Delta & \approx \frac{\sqrt2 v^2 \mu}{2 \mu_{\Delta}^2 +v^2 (\lambda_{H \Delta}+\lambda'_{H \Delta})},  \nonumber  \\
& \approx 0.4 \left( \frac{\rm TeV}{\mu_\Delta} \right)^2 \left( \frac{\mu}{\rm 10 \ GeV} \right) \ [{\rm  GeV}]
\end{align}
where we have assumed $\mu_\Delta^2 \gg v^2 (\lambda_{H\Delta}+\lambda'_{H\Delta})$ at the second line. Combining this result with Eq.~(\ref{eq:mu}), we thus find the small $v_\Delta$ naturally when we take scales of $U(1)_{B-L}$ breaking and triplet mass as $\mathcal{O}(1)$-$\mathcal{O}(10)$ TeV. 
Then the scalar fields are parameterized by
\begin{align}
&H =\left[\begin{array}{c}
w^+\\
\frac{v + h_R +i z}{\sqrt2}
\end{array}\right],\quad 
\Delta =\left[
\begin{array}{cc}
\delta^+/\sqrt2 & \delta^{++}\\
\delta^0 & -\delta^{+}/\sqrt2\\
\end{array}\right],\quad 
\varphi=
\frac{v'+\varphi_R + iz'_\varphi}{\sqrt2},
\label{component}
\end{align}
where $\delta^0\equiv \frac{v_\Delta+\delta_R+iz'_\delta}{\sqrt2}$, $w^+$ and $z$ are absorbed by the SM gauge bosons $W^+$ and $Z$, 
and the massless CP odd boson after diagonalizing the matrix in basis of $(z'_\varphi,z'_\delta)$ is  absorbed by the $B-L$ gauge boson $Z'$.
Although we have mass matrix for CP even scalar bosons in basis of $(\varphi_R,h_R,\delta_R)$, we only consider the matrix as two by two in basis of  $(\varphi_R,h_R)$ ignoring a mixing between $\delta_R$ and other scalars since it is suppressed by small $v_\Delta$. 
The mass matrix for $(\varphi_R,h_R)$ is thus given by
\begin{equation}
M_{\varphi_R h_R} = \begin{pmatrix} 2 \mu_{\varphi}^2 & \lambda_{\varphi H} v v'  \\ \lambda_{\varphi H} v v' & 2 \mu_H^2 \end{pmatrix}. 
\end{equation}
Diagonalizing the matrix, we obtain the following mass eigenvalues:
\begin{equation}
m_{H_1(H_2)} = \mu_\varphi^2 + \mu_H^2 -(+) \sqrt{(\mu_H^2 - \mu_\varphi^2)^2 + (\lambda_{\varphi H} v v')^2}. 
\label{eq:masses}
\end{equation}
Then we parametrize the mixing as
\begin{align}
\left[\begin{array}{c}
\varphi_R \\ h_R
\end{array}\right]
&\equiv
\left[\begin{array}{cc}
c_\theta & -s_\theta \\ 
s_\theta & c_\theta \\
\end{array}\right]
\left[\begin{array}{c}
H_1 \\ H_2
\end{array}\right],\quad t_{2\theta}=\frac{ \lambda_{\varphi H}v v'}{\mu_{H}^2 - \mu_{\varphi}^2},
\label{eq:mixing}
\end{align}
where $H_{1(2)}$ is the mass eigenstate, $t_\theta \equiv \tan \theta$, $c_\theta\equiv \cos \theta$ and $s_\theta\equiv \sin \theta$. In our scenario, $H_1$ is lighter than SM-like Higgs $H_2$~\footnote{We also use the notation $h_{SM}$ for $H_2$ in order to represent the meaning more clearly.}.
The mixing angle is constrained by global analysis for experimental data regarding the SM Higgs production cross section and decay ratio measured by the LHC experiments~\cite{hdecay, Chpoi:2013wga, Cheung:2015dta,Dupuis:2016fda}. 
We then discuss $H_2 \to H_1 H_1$ decay where the relevant interaction is obtained from the potential as follows:  
\begin{align}
-\mathcal{L} & \supset \left[\lambda_H c_\theta s_\theta^2  v - \lambda_\varphi c_\theta^2 s_\theta v' + \frac{1}{4} \lambda_{\varphi H} \{v' s_\theta(2-3 s_\theta^2) + v c_\theta (1-3 s_\theta^2) \} \right] H_2 H_1 H_1 \nonumber \\
& \equiv \frac{\tilde \mu}{2} H_2 H_1 H_1
\end{align}
Applying the interaction, we obtain partial decay width for $H_2 \to H_1 H_1$ such that 
\begin{equation}
\Gamma_{H_2 \to H_1 H_1} = \frac{\tilde \mu^2}{16 \pi m_{H_2}} \sqrt{1- \frac{4 m_{H_1}^2}{m_{H_2}^2}}.
\end{equation}
{Then we estimate the branching ratio for the decay mode of $H_2 \to H_1 H_1$ applying our formulas for VEVs, mass eigenvalues and mixing angle in Eqs.~(\ref{eq:VEVs}), (\ref{eq:masses}) and (\ref{eq:mixing}).
The branching ratio is given by
\begin{equation}
BR(H_2 \to H_1 H_1) = \frac{\Gamma_{H_2 \to H_1 H_1}}{\Gamma_{H_2 \to H_1 H_1} + \Gamma_{h_{SM}}},
\end{equation}
where the decay width of the SM Higgs is given by $\Gamma_{h_{SM}} = 4.2$ MeV. 
The Fig.~\ref{fig:BR} shows the $BR(H_2 \to H_1 H_1)$ as a function of $\sin \theta$ which is compared with current experimental bound for invisible decay branching ratio of the SM Higgs boson~\cite{Olive:2016xmw};
here we take $v'=15$ TeV and $m_{H_1} = 10$ GeV as reference values.
We find that $\sin \theta \lesssim 0.08$ is required to satisfy the constraint.
}

\begin{figure}[t]
\begin{center}
\includegraphics[width=70mm]{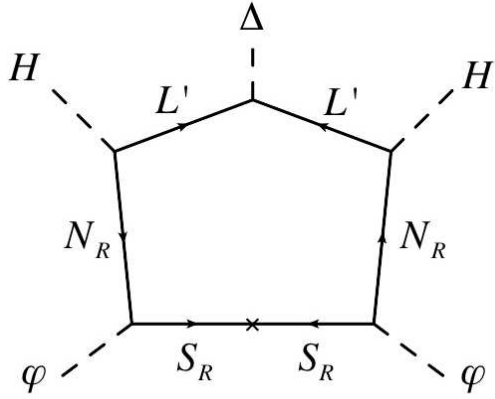} \qquad
\caption{The one loop diagram which induces effective $H^T \Delta H$ term after $U(1)_{B-L}$ symmetry breaking. } 
  \label{fig:VEV}
\end{center}\end{figure}

\begin{figure}[t]
\begin{center}
\includegraphics[width=70mm]{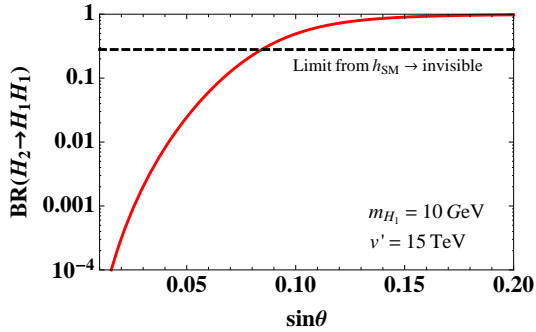} \qquad
\caption{The branching ration for $H_2 \to H_1 H_1$ as a function of $\sin \theta$. } 
  \label{fig:BR}
\end{center}\end{figure}

  \subsection{Lepton sector}
The charged lepton masses are given by $m_\ell =y_\ell v/\sqrt2$ after the electroweak symmetry breaking, where $m_\ell$ is assumed to be the mass eigenstate.
{The neutrino mass matrix arises from $v_\Delta$ at the one-loop level, and the resulting form is given by}
  \begin{align}
({\cal M}_{\nu})_{a b} & = \frac{(y_\nu)_{ab}v_\Delta}{\sqrt2}.
  \end{align}
Then it can reproduce the neutrino oscillation data~\cite{Olive:2016xmw}.
Notice here that this type of model induces the lepton flavor violating processes even at tree level~\cite{Nomura:2017abh}.
The most stringent constraint comes from $\mu\to 3e$ process, and its upper bound is given by
\[
|(y_\nu)_{12}(y_\nu)_{11}^*|\lesssim 2.3\times10^{-5}
\left(\frac{m_{\delta^{\pm\pm}}}{\text{TeV}}\right)^2.
 \]
 Thus we conservatively set the $y_\nu$ to be less than ${\cal O}(10^{-3})$ in order to avoid this constraint.
 In our scenario, the constraints are easily satisfied since we consider $v_\Delta = \mathcal{O}(1)$ GeV and the Yukawa coupling $y_\nu$ can be sufficiently small.

\subsection{ Dark matter} 
At first, let us remind that $\psi_1 \equiv X$ is the DM candidate with mass eigenvalue $m_X$ as we discussed above.
Numerically diagonalizing the mass matrix, we find that the scale of DM mass is $m_X\lesssim {\cal O}(1)$-$\mathcal{O}(10)$ GeV.
This is because the DM mass is given by $m_X\approx m \times \epsilon$ with $m=y_\Delta v_\Delta/\sqrt2\lesssim {\cal O}(10)$ GeV and $\epsilon<1$ is written by complicated combinations of mass parameters in Eq.(\ref{eq-Nmass}). 
Then the main annihilation modes to explain the observed relic density of DM, $\Omega h^2\approx 0.12$~\cite{Ade:2013zuv}, are found to be the SM fermion pairs ($m_f \le m_b$) via s-channel process exchanging $H_1$ and $H_2$.
The relevant Lagrangian in basis of mass eigenstate is given by
\begin{align}
-{\cal L}&= \sum_{a=3}^5\left[
\frac{y_{N_a}}{\sqrt2}(O^*)_{17}(O^T)_{a1} (s_\theta H_1 + c_\theta H_2) +
\sum_{j=1}^2\frac{y_{N_a}}{\sqrt2}(O)_{1a}(O^T)_{j1} (-c_\theta H_1 + s_\theta H_2)\right]
\bar X P_R X \nn\\
&+
\sum_{f\le b}\frac{m_f}{v}\bar f f (s_\theta H_1 + c_\theta H_2)+{\rm c.c.}
\nn\\&
\equiv 
Y_N (s_\theta' H_1 - c_\theta' H_2) \bar X P_R X +\sum_{f}\frac{m_f}{v}\bar f f (s_\theta H_1 + c_\theta H_2)+{\rm c.c.},
\end{align}
where we have reparametrized DM-scalar interactions by $Y_N$ and $s_\theta'(c_\theta')$ for simplicity.
Note that spin independent DM-Nucleon scattering cross section can be calculated by $H_{1,2}$ exchanging diagrams such that~\cite{Okada:2012np} 
\begin{align}\sigma_{ SI} & \approx 8.5\times 10^{-2}\times \frac{\mu_{DM}^2}{2\pi m_{H_1}^4}\left(\frac{Y_N m_N s_\theta^2 }{v}\right)^2  \nonumber \\
& \approx 7.7 \times 10^{-39}\times \left( \frac{\mu_{DM}}{\rm GeV} \right)^2 \left( \frac{\rm 10 \ GeV}{ m_{H_1}} \right)^4 Y_N^2 s_\theta^4 \ [{\rm cm^2}],
 \end{align}
where $\mu_{DM}\equiv \frac{m_N m_X}{m_X+m_N}$, $m_N\approx0.939$ is the neutron mass, the first numerical coefficient in the first line is given by lattice calculation, and we also assume $s_\theta\approx s_\theta'$ for simplicity. We then find that applying typical values in our scenario, {\it e.g.}, the $m_X=10$ GeV, $m_{H_1}=20$ GeV, and ${s_\theta=0.05}$~\footnote{This set gives the stringent constraints of the direct detection searches, and we will use this benchmark point in the analysis of relic density below.}, the upper bound of $Y_N$ is found to be ${\cal O}(0.5)$ in order to satisfy the stringent constraint of the direct detection search at the XENON1T experiment~\cite{Aprile:2017iyp}.
Hereafter $Y_N \lesssim {\cal O}(10^{-1})$ is conservatively imposed. 
Also mixing between SM Higgs and exotic scalar is constrained to be $s_\theta \lesssim 0.08$ as discussed above.
The relic density of DM is then given by~\cite{Griest:1990kh,Edsjo:1997bg}
\begin{align}
&\Omega h^2
\approx 
\frac{1.07\times10^9}{\sqrt{g_*(x_f)}M_{Pl} J(x_f)[{\rm GeV}]},
\label{eq:relic-deff}
\end{align}
where $g^*(x_f\approx25)$ is the degrees of freedom for relativistic particles at temperature $T_f = m_X/x_f$, $M_{Pl}\approx 1.22\times 10^{19}$,
and $J(x_f) (\equiv \int_{x_f}^\infty dx \frac{\langle \sigma v_{\rm rel}\rangle}{x^2})$ is given by~\cite{Nishiwaki:2015iqa}
\begin{align}
J(x_f)&=\int_{x_f}^\infty dx\left[ \frac{\int_{4m_X^2}^\infty ds\sqrt{s-4 m_X^2} s (\sigma v_{\rm rel}) K_1\left(\frac{\sqrt{s}}{m_X} x\right)}{16  m_X^5 x [K_2(x)]^2}\right],\\ 
(\sigma v_{\rm rel})
&\approx\sum_{f\le b}\frac{c_f m_f^2 Y_N^2}{32\pi s v^2}\sqrt{1-\frac{4 m_{f}^2}{s}}
\left(\frac{s_\theta^2}{s-m^2_{H_1}+i m_{H_1} \Gamma_{H_1} } + \frac{c_\theta^2}{s-m^2_{H_2}+i m_{H_2} \Gamma_{H_2}}\right)^2\nn\\
&\times
\left(\frac{s}{2}-m_X^2\right)\left(\frac{s}{2}-2m_f^2\right).
\label{eq:relic-deff}
\end{align}
Here $m_{H_2}=$125 GeV, $\Gamma_{H_2}=$0.0041 GeV, $c_f=3(1)$ for $f$ corresponding to quarks except of top quarks (leptons), $s$ is  the Mandelstam variable, and $K_{1,2}$ are the modified Bessel functions of the second kind of order 1 and 2, respectively.
To analyze the relic density of DM, we fix several values; $Y_N=0.05$, ${s_\theta=0.05}$.
The decay width of $H_1$ is given by
\begin{equation}
\Gamma_{H_1} = \sum_f \frac{m_{H_1}}{8 \pi} \left( \frac{m_f}{v} \right)^2 s_\theta^2 \left( 1 - \frac{4 m_f^2}{m_{H_1}^2} \right)^{\frac{3}{2}}; \qquad (2m_f < m_{H_1}).
\end{equation}
Then we show the relic density in terms of the DM mass in Fig.~\ref{fig:relic} for several values of $m_{H_1} = (5,10,20)$ GeV; note also that we obtain $g^*(x_f) \sim 60$ for $m_{H_1} = (5,10,20)$ GeV for $x_f \simeq 25$.
The results in Fig.~\ref{fig:relic} suggests that each of the solution lies on the pole, $m_{H_1} \simeq 2 m_X$, since resonant enhancement is required due to the tiny coupling.

Here we briefly discuss constraints from CMB power spectrum and reionization history~\cite{Ade:2015xua,Slatyer:2015kla, Liu:2016cnk} since the annihilations of DM into charged particles or photon affect them. Thermal cross section for the DM annihilation is constrained to be $\langle \sigma v \rangle < 10^{-27}-10^{-26}$ cm$^2$ from Planck data~\cite{Ade:2015xua} for CMB spectrum when DM pair dominantly annihilates into $e^+ e^-$ for $m_X \sim \mathcal{O}(1)$-$\mathcal{O}(10)$ GeV. On the other hand, the reionization effect is less significant when the cross section satisfy the Planck constraint~\cite{Liu:2016cnk}. In our scenario, DM pair mainly annihilates into $\bar b b$ or $\tau^+ \tau^-$ depending on $m_X$ and the constraint is less stringent than that from $e^+ e^-$ mode. However the region with $2 m_X  \lesssim m_{H_1}$ would be restricted since $\langle \sigma v \rangle_{\rm after} > \langle \sigma v \rangle_{\rm freeze out}$ ($\langle \sigma v \rangle_{\rm after}$ is thermal cross section after freeze out) due to Breit-Wigner enhancement. On the other hand, the region with $2 m_X  \gtrsim m_{H_1}$ is safe from the constraints since we obtain $\langle \sigma v \rangle_{\rm after} < \langle \sigma v \rangle_{\rm freeze out}$.

Before closing the section, we also comment on the possibility of FIMP scenario in which relic density of DM is explained by freeze-in mechanism~\cite{McDonald:2001vt, Hall:2009bx}.
We can apply FIMP scenario when the couplings for DM-scalar interaction are very small as $Y_N \lesssim \mathcal{O}(10^{-10})$ and DM is out of thermal bath in early Universe.
In such a case, $v_\Delta$ would be too small if we assume $y_N \lesssim \mathcal{O}(10^{-10})$ due to small $\mu$ by Eq.~(\ref{eq:mu}), or we need to fine-tune the mixings among neutral fermions and/or neutral scalar bosons. In this paper, detailed analysis of this scenario is beyond the scope and will be done elsewhere.

\begin{figure}[t]
\centering
\includegraphics[width=10cm]{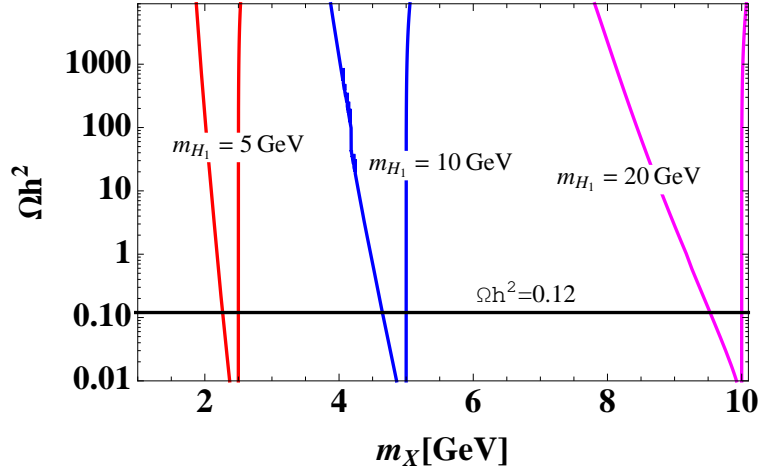}
\caption{Relic density in terms of the DM mass, where $m_{H_1} = (5,10,20)$ GeV represent the lines of red, blue, and magenta, respectively.
Here we fixed $Y_N=0.05$, ${s_\theta=0.05}$, $\Gamma_{H_1}=$0.001 GeV for simplicity. }
\label{fig:relic}
\end{figure}

\subsection{Implications to collider physics}

\begin{figure}[t]
\centering
\includegraphics[width=8cm]{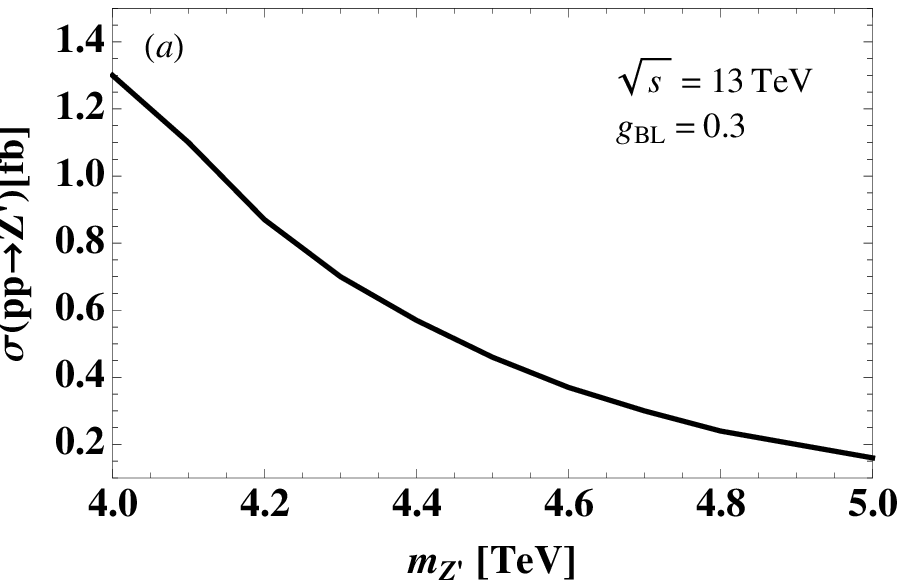}
\includegraphics[width=8cm]{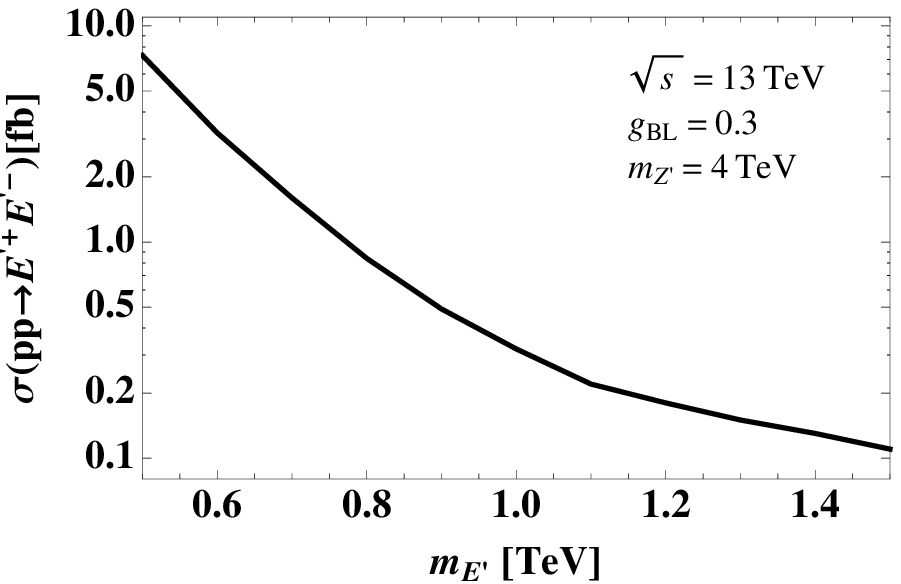}
\caption{(a): The $Z'$ production cross section (in unit of fb) at the LHC 13 TeV as a function of $m_{Z'}$ with $g_{BL}=0.3$. (b): The cross section of $E'^+ E'^-$ pair production (in unit of fb) at the LHC 13 TeV as a function of $m_{E'}$ applying $m_{Z'}=4$ TeV and $g_{BL}=0.3$}
\label{fig:CX}
\end{figure}

In this model, we have Higgs triplet which contain doubly charged Higgs $\delta^{\pm \pm}$ and singly charged Higgs $\delta^\pm$.
For $v_\Delta \sim \mathcal{O}(1)$ GeV, they decay into SM gauge bosons as $\delta^{\pm \pm} \to W^\pm W^\pm$ and $\delta^\pm \to W^\pm Z$.
In particular, doubly charged Higgs provides larger production cross section than singly charged Higgs and gives signal of same sign dilepton (+ missing transverse energy)~\cite{Kanemura:2013vxa, Kanemura:2014ipa,Kanemura:2014goa,Chiang:2012dk}. 
In this case, the mass of the triplet is constrained as $m_\Delta \gtrsim 84$ GeV by the current experimental search for same sign dilepton signal at the LHC
where the doubly charged Higgs is assumed to be produced by electroweak processes~\cite{Kanemura:2014ipa}.
The Higgs triplet can be produced via $Z'$ boson as $pp \to Z' \to \{\delta^{\pm \pm} \delta^{\mp \mp}, \delta^\pm \delta^\mp, \delta^0 \delta^0 \}$ in our model.
Thus It will be interesting to search for signal of $Z' \to \delta^{\pm \pm} \delta^{\mp \mp} (\delta^\pm \delta^\mp) \to W^+ W^+ W^- W^- (W^+ W^- Z Z)$ as a signature of our model where 
produced $\delta^{\pm \pm} (\delta^\pm)$ and resulting $W$ and/or $Z$ bosons in the final states will be highly boosted when $Z'$ mass is much heavier than them.

{We estimate the $Z'$ production cross section at the LHC 13 TeV using {\it CalcHEP}~3.6~\cite{Belyaev:2012qa} by implementing the $Z'$ gauge interactions.
The Fig.~{\ref{fig:CX}}(a) shows the $Z'$ production cross section as a function of $m_{Z'}$ where we applied $g_{BL} =0.3$ as  a reference value.
We find that the cross section is $\sim 1$ fb for $m_{Z'} = 4$ TeV, and the cross section can be scaled as $(g_{BL}/0.3)^2$.
The branching ratios of $Z'$ are also calculated by {\it CalcHEP}~3.6 and we summarize the ratios for some modes of our interest in Table~\ref{tab:BRZp}.
We find that $BR \sim 0.1$ is given for charged scalars, SM leptons and exotic charged lepton pairs where dependence on $m_{Z'}$ is negligible. 
The expected number of events are also shown in Table~\ref{tab:events} for several values of $m_{Z'}$ with the integrated luminosity of 300 fb$^{-1}$. 
Here we just show the potential for discovering $Z'$ signature of our model and the detailed event simulation including SM background with kinematic cuts is beyond the scope of this work which will be given elsewhere.
}

\begin{table}[t]
\centering {\fontsize{10}{12}
\begin{tabular}{c||cccccc}\hline\hline
Modes & ~~$\delta^{++} \delta^{--}$~~ & ~~$\delta^{+} \delta^{-}$~~ & ~~$\ell^+ \ell^-$~~ & ~~$jj$~~ & ~~$t \bar t$~~ & ~~$E'^+ E'^-$~~ \\ \hline
BRs & 0.088 & 0.088 & 0.089 & 0.12 & 0.030 & 0.081 \\ \hline
\end{tabular}%
} 
\caption{The branching ratio of $Z'$ for some modes where $\ell = \{e, \mu, \tau \}$ and $m_{E'} = 1$ TeV is used. }
\label{tab:BRZp}
\end{table}

\begin{table}[t]
\centering {\fontsize{10}{12}
\begin{tabular}{c||cccccc}\hline\hline
Modes & ~~$\delta^{++} \delta^{--}$~~ & ~~$\delta^{+} \delta^{-}$~~ & ~~$\ell^+ \ell^-$~~ & ~~$jj$~~ & ~~$t \bar t$~~ & ~~$E'^+ E'^-$~~ \\ \hline
$m_{Z'}$ = 4 TeV & 34 & 34 & 35 & 47 & 12 & 32 \\ 
 4.5 TeV & 12 & 12 & 12 & 17 & 4 & 11 \\ 
  5.0 TeV & 4 & 4 & 4 & 6 & 1 & 4 \\ \hline
\end{tabular}%
} 
\caption{The expected number of events for $pp \to Z' \to F \bar F$ for some values of $m_{Z'}$ where $F$ indicates each final states and the integrated luminosity is taken to be 300 fb$^{-1}$. }
\label{tab:events}
\end{table}

{
The exotic charged lepton $E'^\pm$ can also be produced by electroweak interaction since it is from vector like exotic lepton doublet $L'$.
Here we estimate the $E'^\pm$ production cross section including electroweak and $Z'$ interactions. 
The Fig.~\ref{fig:CX}(b) shows the cross section as a function of $m_{E'}$ where we have applied $g_{BL} =0.3$ and $m_{Z'} = 4$ TeV as reference values.
The $E'^-$ dominantly decays as $E'^- \to \delta^- X$ via Yukawa interaction.
Thus, from $E'^+ E'^-$ pair production at the LHC, we obtain signal of $W^+ W^- Z Z + \slashed{E}_T$ where gauge bosons are produced via decay of $\delta^\pm$.
These signals can be tested in future experiment at the LHC with sufficient integrated luminosity. 
For example, the cross section of $E'^+E'^-$ pair production is estimated as $\sigma_{pp\to E'^+ E'^-} \simeq 0.4$ fb when the mass of $E'^-$ is 1 TeV; electroweak production is dominant and around $0.1$ fb is obtained for $pp \to Z' \to E'^+ E'^-$.  Thus roughly $100$ events are expected for $E'^+E'^-$ pair with integrated luminosity of 300 fb$^{-1}$.
As the $Z'$ production case, here we leave the detailed event simulation study in future work.
}

\section{ Conclusions and discussions}
We have propose a model with $U(1)_{B-L}$ gauge symmetry and several new fermions in no conflict with anomaly cancellation.
Then the neutrino masses are given by VEV of Higgs triplet induced at one-loop level, and
the new fermions are odd under discrete $Z_2$ symmetry and the lightest one becomes dark matter candidate.

We have shown the mechanism to generate the neutrino masses  and have analyzed a fermionic dark matter candidate.
The scale of DM mass is found to be ${\cal O}(1-10)$ GeV by the structure of the neutral fermion mass matrices.
Then, we have explicitly shown several solutions to satisfy the relic density of DM without conflict of direct detection experiment, fixing $Y_N = s_\theta = 0.05$ and $\Gamma_{H_1} = 0.001$ with $m_{H_1} = (5,10,20)$ GeV. Our analysis has suggested that each of the solution lies on the pole, $m_{H_1} \simeq 2 m_X$, due to the tiny coupling.
In addition, we find too small mass of $H_1$ is not allowed by relic density even when it is on the pole. 
We have also discussed implications to collider physics where the $Z'$ and exotic charged lepton pair production cross sections are estimated.
Then the potential for discovering signature of our model has been indicated showing expected number of events at the LHC 13 TeV where detailed event simulation study including SM background estimation with kinematical cuts is beyond the scope of this paper and it is left as future work.

\section*{Acknowledgments}
\vspace{0.5cm}
H. O. is sincerely grateful for all the KIAS members, Korean cordial persons, foods, culture, weather, and all the other things.


\begin{thebibliography}{99}

\bibitem{a-zee} 
A.~Zee,
 Phys.\ Lett.\  B {\bf 93}, 389 (1980)
 [Erratum-ibid.\  B {\bf 95}, 461  (1980)].

\bibitem{Cheng-Li} 
  T.~P.~Cheng and L.~F.~Li,
  Phys.\ Rev.\ D {\bf 22}, 2860 (1980).

 \bibitem{Pilaftsis:1991ug} 
  A.~Pilaftsis,
  Z.\ Phys.\ C {\bf 55}, 275 (1992)
  [hep-ph/9901206].

\bibitem{Ma:2006km} 
  E.~Ma,
  Phys.\ Rev.\ D {\bf 73}, 077301 (2006)
  [hep-ph/0601225].

\bibitem{Gu:2007ug} 
  P.~-H.~Gu and U.~Sarkar,
  Phys.\ Rev.\ D {\bf 77}, 105031 (2008)
  [arXiv:0712.2933 [hep-ph]].

\bibitem{Sahu:2008aw} 
  N.~Sahu and U.~Sarkar;
  Phys.\ Rev.\ D {\bf 78}, 115013 (2008)
  [arXiv:0804.2072 [hep-ph]].
  
\bibitem{Gu:2008zf} 
  P.~-H.~Gu and U.~Sarkar,
  Phys.\ Rev.\ D {\bf 78}, 073012 (2008)
  [arXiv:0807.0270 [hep-ph]].
  
\bibitem{AristizabalSierra:2006ri} 
  D.~Aristizabal Sierra and D.~Restrepo,
  JHEP {\bf 0608}, 036 (2006)
  [hep-ph/0604012].
  
\bibitem{Bouchand:2012dx} 
  R.~Bouchand and A.~Merle,
  JHEP {\bf 1207}, 084 (2012)
  [arXiv:1205.0008 [hep-ph]].
  
\bibitem{McDonald:2013hsa} 
  K.~L.~McDonald,
  JHEP {\bf 1311}, 131 (2013)
  [arXiv:1310.0609 [hep-ph]].
  
\bibitem{Ma:2014cfa} 
  E.~Ma,
  Phys.\ Lett.\ B {\bf 732}, 167 (2014)
  [arXiv:1401.3284 [hep-ph]].

\bibitem{Kajiyama:2013sza} 
  Y.~Kajiyama, H.~Okada and K.~Yagyu,
  Nucl.\ Phys.\ B {\bf 887}, 358 (2014)
  [arXiv:1309.6234 [hep-ph]].

\bibitem{Kanemura:2011vm} 
  S.~Kanemura, O.~Seto and T.~Shimomura,
  Phys.\ Rev.\ D {\bf 84}, 016004 (2011)
  [arXiv:1101.5713 [hep-ph]].

\bibitem{Kanemura:2011jj}
 S.~Kanemura, T.~Nabeshima and H.~Sugiyama,
 Phys.\ Lett.\ B {\bf 703}, 66 (2011)
 [arXiv:1106.2480 [hep-ph]].

\bibitem{Kanemura:2011mw} 
  S.~Kanemura, T.~Nabeshima and H.~Sugiyama,
  Phys.\ Rev.\ D {\bf 85}, 033004 (2012)
  [arXiv:1111.0599 [hep-ph]].

\bibitem{Schmidt:2012yg} 
  D.~Schmidt, T.~Schwetz and T.~Toma,
  Phys.\ Rev.\ D {\bf 85}, 073009 (2012)
  [arXiv:1201.0906 [hep-ph]].

\bibitem{Kanemura:2012rj} 
  S.~Kanemura and H.~Sugiyama,
  Phys.\ Rev.\ D {\bf 86}, 073006 (2012)
  [arXiv:1202.5231 [hep-ph]].

\bibitem{Farzan:2012sa} 
  Y.~Farzan and E.~Ma,
  Phys.\ Rev.\ D {\bf 86}, 033007 (2012)
  [arXiv:1204.4890 [hep-ph]].

\bibitem{Kumericki:2012bf} 
  K.~Kumericki, I.~Picek and B.~Radovcic,
  JHEP {\bf 1207}, 039 (2012)
  [arXiv:1204.6597 [hep-ph]].

\bibitem{Kumericki:2012bh} 
  K.~Kumericki, I.~Picek and B.~Radovcic,
  Phys.\ Rev.\ D {\bf 86}, 013006 (2012)
  [arXiv:1204.6599 [hep-ph]].

\bibitem{Ma:2012if} 
  E.~Ma,
  Phys.\ Lett.\ B {\bf 717}, 235 (2012)
  [arXiv:1206.1812 [hep-ph]].

\bibitem{Gil:2012ya} 
  G.~Gil, P.~Chankowski and M.~Krawczyk,
  Phys.\ Lett.\ B {\bf 717}, 396 (2012)
  [arXiv:1207.0084 [hep-ph]].

\bibitem{Okada:2012np} 
  H.~Okada and T.~Toma,
  Phys.\ Rev.\ D {\bf 86}, 033011 (2012)
  arXiv:1207.0864 [hep-ph].

\bibitem{Hehn:2012kz} 
  D.~Hehn and A.~Ibarra,
  Phys.\ Lett.\ B {\bf 718}, 988 (2013)
  [arXiv:1208.3162 [hep-ph]].

\bibitem{Dev:2012sg} 
  P.~S.~B.~Dev and A.~Pilaftsis,
  Phys.\ Rev.\ D {\bf 86}, 113001 (2012)
  [arXiv:1209.4051 [hep-ph]].

\bibitem{Kajiyama:2012xg} 
  Y.~Kajiyama, H.~Okada and T.~Toma,
  Eur.\ Phys.\ J.\ C {\bf 73}, 2381 (2013)
  [arXiv:1210.2305 [hep-ph]].

\bibitem{Toma:2013zsa} 
  T.~Toma and A.~Vicente,
  JHEP {\bf 1401}, 160 (2014)
  [arXiv:1312.2840, arXiv:1312.2840 [hep-ph]].

\bibitem{Kanemura:2013qva} 
  S.~Kanemura, T.~Matsui and H.~Sugiyama,
  Phys.\ Lett.\ B {\bf 727}, 151 (2013)
  [arXiv:1305.4521 [hep-ph]].

\bibitem{Law:2013saa} 
  S.~S.~C.~Law and K.~L.~McDonald,
  JHEP {\bf 1309}, 092 (2013)
  [arXiv:1305.6467 [hep-ph]].

\bibitem{Baek:2014qwa} 
  S.~Baek and H.~Okada,
  arXiv:1403.1710 [hep-ph].

\bibitem{Kanemura:2014rpa} 
  S.~Kanemura, T.~Matsui and H.~Sugiyama,
  Phys.\ Rev.\ D {\bf 90}, 013001 (2014)
  [arXiv:1405.1935 [hep-ph]].

\bibitem{Fraser:2014yha} 
  S.~Fraser, E.~Ma and O.~Popov,
  Phys.\ Lett.\ B {\bf 737}, 280 (2014)
  [arXiv:1408.4785 [hep-ph]].

\bibitem{Vicente:2014wga} 
  A.~Vicente and C.~E.~Yaguna,
  JHEP {\bf 1502}, 144 (2015)
  [arXiv:1412.2545 [hep-ph]].

\bibitem{Baek:2015mna} 
  S.~Baek, H.~Okada and K.~Yagyu,
  JHEP {\bf 1504}, 049 (2015)
  [arXiv:1501.01530 [hep-ph]].

\bibitem{Merle:2015gea} 
  A.~Merle and M.~Platscher,
  Phys.\ Rev.\ D {\bf 92}, no. 9, 095002 (2015)
  [arXiv:1502.03098 [hep-ph]].
  
\bibitem{Restrepo:2015ura} 
  D.~Restrepo, A.~Rivera, M.~S\'anchez-Pel\'aez, O.~Zapata and W.~Tangarife,
  Phys.\ Rev.\ D {\bf 92}, no. 1, 013005 (2015)
  [arXiv:1504.07892 [hep-ph]].

\bibitem{Merle:2015ica} 
  A.~Merle and M.~Platscher,
  JHEP {\bf 1511}, 148 (2015)
  [arXiv:1507.06314 [hep-ph]].

\bibitem{Wang:2015saa} 
  W.~Wang and Z.~L.~Han,
  Phys.\ Rev.\ D {\bf 92}, 095001 (2015)
  [arXiv:1508.00706 [hep-ph]].

 \bibitem{Ahn:2012cg} 
  Y.~H.~Ahn and H.~Okada,
  Phys.\ Rev.\ D {\bf 85}, 073010 (2012)
  [arXiv:1201.4436 [hep-ph]].

\bibitem{Ma:2012ez} 
  E.~Ma, A.~Natale and A.~Rashed,
  Int.\ J.\ Mod.\ Phys.\ A {\bf 27}, 1250134 (2012)
  [arXiv:1206.1570 [hep-ph]].

\bibitem{Hernandez:2013dta} 
  A.~E.~Carcamo Hernandez, I.~d.~M.~Varzielas, S.~G.~Kovalenko, H.~P\"{a}s and I.~Schmidt,
  Phys.\ Rev.\ D {\bf 88}, 076014 (2013)
  [arXiv:1307.6499 [hep-ph]].

\bibitem{Ma:2014eka} 
  E.~Ma and A.~Natale,
  Phys.\ Lett.\ B {\bf 723}, 403 (2014)
  [arXiv:1403.6772 [hep-ph]].

\bibitem{Ma:2014yka} 
  E.~Ma,
  Phys.\ Lett.\ B {\bf 741}, 202 (2015)
  [arXiv:1411.6679 [hep-ph]].

\bibitem{Ma:2015pma} 
  E.~Ma,
  Phys.\ Rev.\ D {\bf 92}, no. 5, 051301 (2015)
  [arXiv:1504.02086 [hep-ph]].

\bibitem{Ma:2013mga} 
  E.~Ma,
  Phys.\ Rev.\ Lett.\  {\bf 112}, 091801 (2014)
  [arXiv:1311.3213 [hep-ph]].

\bibitem{radlepton1} 
  H.~Okada and K.~Yagyu,
  Phys.\ Rev.\ D {\bf 89}, 053008 (2014)
  [arXiv:1311.4360 [hep-ph]].

\bibitem{Okada:2014nsa} 
  H.~Okada and K.~Yagyu;
  Phys.\ Rev.\ D {\bf 90}, no. 3, 035019 (2014)
  [arXiv:1405.2368 [hep-ph]].

\bibitem{Brdar:2013iea} 
  V.~Brdar, I.~Picek and B.~Radovcic,
  Phys.\ Lett.\ B {\bf 728}, 198 (2014)
  [arXiv:1310.3183 [hep-ph]].

\bibitem{Okada:2015kkj} 
  H.~Okada, Y.~Orikasa and T.~Toma,
  Phys.\ Rev.\ D {\bf 93}, no. 5, 055007 (2016)
  [arXiv:1511.01018 [hep-ph]].

\bibitem{Bonnet:2012kz} 
  F.~Bonnet, M.~Hirsch, T.~Ota and W.~Winter,
  JHEP {\bf 1207}, 153 (2012)
  [arXiv:1204.5862 [hep-ph]].

\bibitem{Joaquim:2014gba} 
  F.~R.~Joaquim and J.~T.~Penedo,
  Phys.\ Rev.\ D {\bf 90}, no. 3, 033011 (2014)
  [arXiv:1403.4925 [hep-ph]].

\bibitem{Davoudiasl:2014pya} 
  H.~Davoudiasl and I.~M.~Lewis,
  Phys.\ Rev.\ D {\bf 90}, no. 3, 033003 (2014)
  [arXiv:1404.6260 [hep-ph]].

\bibitem{Lindner:2014oea} 
  M.~Lindner, S.~Schmidt and J.~Smirnov,
  JHEP {\bf 1410}, 177 (2014)
  [arXiv:1405.6204 [hep-ph]].


\bibitem{Okada:2014nea} 
  H.~Okada and Y.~Orikasa,
  Phys.\ Lett.\ B {\bf 760}, 558 (2016)
  [arXiv:1412.3616 [hep-ph]].


\bibitem{Mambrini:2015sia} 
  Y.~Mambrini, S.~Profumo and F.~S.~Queiroz,
  Phys.\ Lett.\ B {\bf 760}, 807 (2016)
  [arXiv:1508.06635 [hep-ph]].
  
  
  
\bibitem{Boucenna:2014zba} 
  S.~M.~Boucenna, S.~Morisi and J.~W.~F.~Valle,
  Adv.\ High Energy Phys.\  {\bf 2014}, 831598 (2014)
  [arXiv:1404.3751 [hep-ph]].

\bibitem{Ahriche:2016acx} 
  A.~Ahriche, S.~M.~Boucenna and S.~Nasri,
  Phys.\ Rev.\ D {\bf 93}, no. 7, 075036 (2016)
  [arXiv:1601.04336 [hep-ph]].

\bibitem{Fraser:2015mhb} 
  S.~Fraser, C.~Kownacki, E.~Ma and O.~Popov,
  Phys.\ Rev.\ D {\bf 93}, no. 1, 013021 (2016)
  [arXiv:1511.06375 [hep-ph]].

\bibitem{Fraser:2015zed} 
  S.~Fraser, E.~Ma and M.~Zakeri,
  Phys.\ Rev.\ D {\bf 93}, no. 11, 115019 (2016)
  [arXiv:1511.07458 [hep-ph]].

\bibitem{Adhikari:2015woo} 
  R.~Adhikari, D.~Borah and E.~Ma,
  Phys.\ Lett.\ B {\bf 755}, 414 (2016)
  [arXiv:1512.05491 [hep-ph]].

\bibitem{Okada:2015vwh} 
  H.~Okada and Y.~Orikasa,
  Phys.\ Rev.\ D {\bf 94}, no. 5, 055002 (2016)
  [arXiv:1512.06687 [hep-ph]].
  
  
  
\bibitem{Ibarra:2016dlb} 
  A.~Ibarra, C.~E.~Yaguna and O.~Zapata,
  Phys.\ Rev.\ D {\bf 93}, no. 3, 035012 (2016)
  [arXiv:1601.01163 [hep-ph]].

\bibitem{Arbelaez:2016mhg} 
  C.~Arbelaez, A.~E.~C.~Hernandez, S.~Kovalenko and I.~Schmidt,
  arXiv:1602.03607 [hep-ph].

\bibitem{Ahriche:2016rgf} 
  A.~Ahriche, K.~L.~McDonald, S.~Nasri and I.~Picek,
  Phys.\ Lett.\ B {\bf 757}, 399 (2016)
  [arXiv:1603.01247 [hep-ph]].

\bibitem{Lu:2016ucn} 
  W.~B.~Lu and P.~H.~Gu,
  JCAP {\bf 1605}, no. 05, 040 (2016)
  [arXiv:1603.05074 [hep-ph]].

\bibitem{Kownacki:2016hpm} 
  C.~Kownacki and E.~Ma,
  Phys.\ Lett.\ B {\bf 760}, 59 (2016)
  [arXiv:1604.01148 [hep-ph]].

\bibitem{Ahriche:2016cio} 
  A.~Ahriche, K.~L.~McDonald and S.~Nasri,
  JHEP {\bf 1606}, 182 (2016)
  [arXiv:1604.05569 [hep-ph]].

\bibitem{Ahriche:2016ixu} 
  A.~Ahriche, A.~Manning, K.~L.~McDonald and S.~Nasri,
  Phys.\ Rev.\ D {\bf 94}, no. 5, 053005 (2016)
  [arXiv:1604.05995 [hep-ph]].

\bibitem{Ma:2016nnn} 
  E.~Ma, N.~Pollard, O.~Popov and M.~Zakeri,
  Mod.\ Phys.\ Lett.\ A {\bf 31}, no. 27, 1650163 (2016)
  [arXiv:1605.00991 [hep-ph]].

\bibitem{Nomura:2016jnl} 
  T.~Nomura, H.~Okada and Y.~Orikasa,
  Phys.\ Rev.\ D {\bf 94}, no. 5, 055012 (2016)
  [arXiv:1605.02601 [hep-ph]].

 
\bibitem{Hagedorn:2016dze} 
  C.~Hagedorn, T.~Ohlsson, S.~Riad and M.~A.~Schmidt,
  JHEP {\bf 1609}, 111 (2016)
  [arXiv:1605.03986 [hep-ph]].
  

\bibitem{Antipin:2016awv} 
  O.~Antipin, P.~Culjak, K.~Kumericki and I.~Picek,
  arXiv:1606.05163 [hep-ph].


  \bibitem{Nomura:2016emz} 
  T.~Nomura and H.~Okada,
  Phys.\ Lett.\ B {\bf 761}, 190 (2016)
  [arXiv:1606.09055 [hep-ph]].
 
\bibitem{Gu:2016ghu} 
  P.~H.~Gu, E.~Ma and U.~Sarkar,
  Phys.\ Rev.\ D {\bf 94}, no. 11, 111701 (2016)
  [arXiv:1608.02118 [hep-ph]].

\bibitem{Guo:2016dzl} 
  S.~Y.~Guo, Z.~L.~Han and Y.~Liao,
  Phys.\ Rev.\ D {\bf 94}, no. 11, 115014 (2016)
  [arXiv:1609.01018 [hep-ph]].

\bibitem{Hernandez:2015hrt} 
  A.~E.~Carcamo Hernandez,
  Eur.\ Phys.\ J.\ C {\bf 76}, no. 9, 503 (2016)
  [arXiv:1512.09092 [hep-ph]].

\bibitem{Megrelidze:2016fcs} 
  L.~Megrelidze and Z.~Tavartkiladze,
  Nucl.\ Phys.\ B {\bf 914}, 553 (2017)
  [arXiv:1609.07344 [hep-ph]].

\bibitem{Cheung:2016fjo} 
  K.~Cheung, T.~Nomura and H.~Okada,
  Phys.\ Rev.\ D {\bf 94}, no. 11, 115024 (2016)
  [arXiv:1610.02322 [hep-ph]].


  
  
\bibitem{Seto:2016pks} 
  O.~Seto and T.~Shimomura,
  arXiv:1610.08112 [hep-ph].
 
\bibitem{Lu:2016dbc} 
  W.~B.~Lu and P.~H.~Gu,
  arXiv:1611.02106 [hep-ph].
 

\bibitem{Hessler:2016kwm} 
  A.~G.~Hessler, A.~Ibarra, E.~Molinaro and S.~Vogl,
  JHEP {\bf 1701}, 100 (2017)
  [arXiv:1611.09540 [hep-ph]].

\bibitem{Okada:2015bxa} 
  H.~Okada, N.~Okada and Y.~Orikasa,
  Phys.\ Rev.\ D {\bf 93}, no. 7, 073006 (2016)
  [arXiv:1504.01204 [hep-ph]].
 
\bibitem{Ko:2017quv} 
  P.~Ko, T.~Nomura and H.~Okada,
  arXiv:1701.05788 [hep-ph].
 
\bibitem{Ko:2017yrd} 
  P.~Ko, T.~Nomura and H.~Okada,
  arXiv:1702.02699 [hep-ph].
 
\bibitem{Lee:2017ekw} 
  S.~Lee, T.~Nomura and H.~Okada,
  arXiv:1702.03733 [hep-ph].
  
   
\bibitem{Antipin:2017wiz} 
  O.~Antipin, P.~Culjak, K.~Kumericki and I.~Picek,
  Phys.\ Lett.\ B {\bf 768}, 330 (2017)
  [arXiv:1703.05075 [hep-ph]].
 
\bibitem{Borah:2017dqx} 
  D.~Borah, S.~Sadhukhan and S.~Sahoo,
  arXiv:1703.08674 [hep-ph].
 
\bibitem{Chiang:2017tai} 
  C.~W.~Chiang, H.~Okada and E.~Senaha,
  arXiv:1703.09153 [hep-ph].
 
\bibitem{Kitabayashi:2017sjz} 
  T.~Kitabayashi, S.~Ohkawa and M.~Yasue,
  arXiv:1703.09417 [hep-ph].
  
  
\bibitem{Das:2017ski} 
  A.~Das, T.~Nomura, H.~Okada and S.~Roy,
  arXiv:1704.02078 [hep-ph].
  
\bibitem{Nomura:2017ezy} 
  T.~Nomura and H.~Okada,
  arXiv:1704.03382 [hep-ph].
  
  




\bibitem{Wang:2016lve} 
  W.~Wang and Z.~L.~Han,
  arXiv:1611.03240 [hep-ph].
 
\bibitem{Guo:2017ybk} 
  C.~Guo, S.~Y.~Guo, Z.~L.~Han, B.~Li and Y.~Liao,
  arXiv:1701.02463 [hep-ph].

\bibitem{Lindner:2016bgg} 
  M.~Lindner, M.~Platscher and F.~S.~Queiroz,
  arXiv:1610.06587 [hep-ph].
  
  
  
\bibitem{Magg:1980ut} 
  M.~Magg and C.~Wetterich,
  Phys.\ Lett.\ B {\bf 94}, 61 (1980);
  G.~Lazarides, Q.~Shafi and C.~Wetterich,
  Nucl.\ Phys.\ B {\bf 181}, 287 (1981);
  R.~N.~Mohapatra and G.~Senjanovic,
  Phys.\ Rev.\ D {\bf 23}, 165 (1981);
  E.~Ma and U.~Sarkar,
  Phys.\ Rev.\ Lett.\  {\bf 80}, 5716 (1998).
 \bibitem{Konetschny:1977bn} 
  W.~Konetschny and W.~Kummer,
  Phys.\ Lett.\ B {\bf 70}, 433 (1977);
  J.~Schechter and J.~W.~F.~Valle,
  Phys.\ Rev.\ D {\bf 22}, 2227 (1980);
  T.~P.~Cheng and L.~-F.~Li,
  Phys.\ Rev.\ D {\bf 22}, 2860 (1980);
  S.~M.~Bilenky, J.~Hosek and S.~T.~Petcov,
  Phys.\ Lett.\ B {\bf 94}, 495 (1980).

  
  
\bibitem{Okada:2015nca} 
  H.~Okada and Y.~Orikasa,
  Phys.\ Rev.\ D {\bf 93}, no. 1, 013008 (2016)
  [arXiv:1509.04068 [hep-ph]].





\bibitem{Olive:2016xmw} 
  C.~Patrignani {\it et al.} [Particle Data Group],
  Chin.\ Phys.\ C {\bf 40}, no. 10, 100001 (2016).
  doi:10.1088/1674-1137/40/10/100001

\bibitem{Nomura:2017abh} 
  T.~Nomura, H.~Okada and H.~Yokoya,
  arXiv:1702.03396 [hep-ph].
  
  
    




  
\bibitem{Ade:2013zuv} 
  P.~A.~R.~Ade {\it et al.} [Planck Collaboration],
  Astron.\ Astrophys.\  {\bf 571}, A16 (2014)
  [arXiv:1303.5076 [astro-ph.CO]].
  
\bibitem{Aprile:2017iyp} 
  E.~Aprile {\it et al.} [XENON Collaboration],
  arXiv:1705.06655 [astro-ph.CO].
  
    
  
    \bibitem{hdecay}  The numerical analyses on the Higgs decays are performed using the program {\tt HDECAY}: A.~Djouadi, J.~Kalinowski and M.~Spira, Comput. Phys.
Commun. 108 (1998) 56; A. Djouadi, M. Muhlleitner and M. Spira, Acta. Phys. Polon. 
B38 (2007) 635. 
  
  
\bibitem{Chpoi:2013wga} 
  S.~Choi, S.~Jung and P.~Ko,
  JHEP {\bf 1310}, 225 (2013)
  [arXiv:1307.3948 [hep-ph]].
 
\bibitem{Cheung:2015dta} 
  K.~Cheung, P.~Ko, J.~S.~Lee and P.~Y.~Tseng,
  JHEP {\bf 1510}, 057 (2015)
  [arXiv:1507.06158 [hep-ph]].

  
\bibitem{Dupuis:2016fda} 
  G.~Dupuis,
  JHEP {\bf 1607}, 008 (2016)
  [arXiv:1604.04552 [hep-ph]].




\bibitem{Griest:1990kh} 
  K.~Griest and D.~Seckel,
  Phys.\ Rev.\ D {\bf 43}, 3191 (1991).

\bibitem{Edsjo:1997bg} 
  J.~Edsjo and P.~Gondolo,
  Phys.\ Rev.\ D {\bf 56}, 1879 (1997)
  [hep-ph/9704361].
  


\bibitem{Nishiwaki:2015iqa} 
  K.~Nishiwaki, H.~Okada and Y.~Orikasa,
  Phys.\ Rev.\ D {\bf 92}, no. 9, 093013 (2015)
  [arXiv:1507.02412 [hep-ph]].
  
\bibitem{Ade:2015xua} 
  P.~A.~R.~Ade {\it et al.} [Planck Collaboration],
  Astron.\ Astrophys.\  {\bf 594}, A13 (2016)
  [arXiv:1502.01589 [astro-ph.CO]].
  
\bibitem{Slatyer:2015kla} 
  T.~R.~Slatyer,
  Phys.\ Rev.\ D {\bf 93}, no. 2, 023521 (2016)
  [arXiv:1506.03812 [astro-ph.CO]].
  
\bibitem{Liu:2016cnk} 
  H.~Liu, T.~R.~Slatyer and J.~Zavala,
  Phys.\ Rev.\ D {\bf 94}, no. 6, 063507 (2016)
  [arXiv:1604.02457 [astro-ph.CO]].
  
\bibitem{McDonald:2001vt} 
  J.~McDonald,
  Phys.\ Rev.\ Lett.\  {\bf 88}, 091304 (2002)
  [hep-ph/0106249].
  
\bibitem{Hall:2009bx} 
  L.~J.~Hall, K.~Jedamzik, J.~March-Russell and S.~M.~West,
  JHEP {\bf 1003}, 080 (2010)
  [arXiv:0911.1120 [hep-ph]].
  
  
\bibitem{Chiang:2012dk} 
  C.~W.~Chiang, T.~Nomura and K.~Tsumura,
  Phys.\ Rev.\ D {\bf 85}, 095023 (2012)
  [arXiv:1202.2014 [hep-ph]].
  
\bibitem{Kanemura:2013vxa} 
  S.~Kanemura, K.~Yagyu and H.~Yokoya,
  Phys.\ Lett.\ B {\bf 726}, 316 (2013)
  [arXiv:1305.2383 [hep-ph]].
  
\bibitem{Kanemura:2014goa} 
  S.~Kanemura, M.~Kikuchi, K.~Yagyu and H.~Yokoya,
  Phys.\ Rev.\ D {\bf 90}, no. 11, 115018 (2014)
  [arXiv:1407.6547 [hep-ph]].

\bibitem{Kanemura:2014ipa} 
  S.~Kanemura, M.~Kikuchi, H.~Yokoya and K.~Yagyu,
  PTEP {\bf 2015}, 051B02 (2015)
  [arXiv:1412.7603 [hep-ph]].
  
    \bibitem{Belyaev:2012qa} 
  A.~Belyaev, N.~D.~Christensen and A.~Pukhov,
  Comput.\ Phys.\ Commun.\  {\bf 184}, 1729 (2013)
  [arXiv:1207.6082 [hep-ph]].

  
\end{thebibliography}
\end{document}